 \documentclass[preprint,10pt]{aastex}

\slugcomment{To appear in the Astronomical Journal}

\shortauthors{Bersier \& Wood}
\shorttitle{Variables stars in Fornax}

\begin{document}


\title{Variable Stars in the Fornax Dwarf Galaxy}


\author{D. Bersier}
\affil{Harvard-Smithsonian Center for Astrophysics, 60 Garden St,
Cambridge MA 02138}

\author{P. R. Wood}
\affil{Research School of Astronomy \& Astrophysics,
Australian National University,
Mount Stromlo Observatory, Private Bag, Weston Creek PO, ACT 2611, Australia}


\begin{abstract}

We present a search for variable stars in the Fornax dwarf galaxy
covering an area of 1/2 a square degree. We have $\sim 30$ epochs of
$VI$ data. We found and determined periods for more than 500 RR Lyrae,
17 anomalous Cepheids, 6 Population II Cepheids.  In addition we have
85 candidate Long Period Variables, the majority of which were
previously unknown.  We estimated that the average metal abundance of
RR Lyrae stars is $\mbox{[Fe/H]} \simeq -1.6$ dex.

\end{abstract}


\keywords{galaxies: individual: Fornax --- stars: RR Lyrae --- distance scale
--- stars: variable}

\section{Introduction}
\label{sec_intro}

Several studies have found variable stars in Fornax or have detected variability
(Light et al. 1986, Demers \& Irwin 1987, Buonanno et al. 1985, Stetson et al.
1998). However they all suffered from some limitations. For instance \citet{shsh}
cover a large area with deep photometry but they don't have enough epochs to determine
periods for their candidate variables. \citet{laz86} did have a lot of data but
they covered a very small area; they have light curves for only two
variable stars. \citet{di87} covered the whole galaxy and had
$\sim 20$ epochs but their photometry didn't go deep enough to detect RR Lyrae.
For these reasons, there are few variable stars known in Fornax with published
periods and light curves.
We thus thought that it would be worthwhile to survey a fair fraction
of this galaxy and look for variable stars, in particular RR Lyrae stars.

\section{Observations, data reduction and photometry}
\label{sec_obs}

Imaging observations of Fornax were made with the 1m telescope at Siding Spring
Observatory and with the 1.3m telescope at Mount Stromlo Observatory.  Roughly
half the epochs of observation came from each telescope.  Table~\ref{tbl_dates}
lists the dates of observation for each field. The two sets of
observations are described separately below.

\subsection{Observations at Siding Spring Observatory}

The part of the data obtained with the 1m telescope at Siding Spring
Observatory was taken during two runs of a few consecutive nights. We used a
SITe $2048 \times 2048$ CCD that gives a field of view of
$20.5' \times 20.5'$ (0.602 arcsecond per pixel). Four fields have been observed
in $V$ and $I_C$ filters, covering a square on the sky.
The four fields were centered approximately on
$\alpha_1 = 2^h 40^m 26.0^s,\ \delta_1 = -34^\circ 28' 40''$,
$\alpha_2 = 2^h 38^m 40.0^s,\ \delta_2 = -34^\circ 28' 40''$,
$\alpha_3 = 2^h 40^m 26.0^s,\ \delta_3 = -34^\circ 50' 00''$,
$\alpha_4 = 2^h 38^m 40.0^s,\ \delta_4 = -34^\circ 50' 00''$
(thus field 1 is north-east, field 2 is north-west, field 3 is
south-east and field 4 is south-west).  The total area observed is
almost half a square degree, covering the central regions of the
galaxy.  Each field has been observed between one and four times per
night, with a typical exposure time of 10 minutes, resulting in about
12 to 15 exposures per field. Our average seeing is 2'', with the best
value being 1.5''.  During two nights we also observed standard stars
taken from the list of \cite{l92}.

The data have been processed in a standard way using IRAF\footnote{
IRAF is distributed by the National Optical Astronomy Observatories,
which are operated by the Association of Universities for Research
in Astronomy, Inc., under cooperative agreement with the National
Science Foundation.} procedures.
We performed aperture photometry for the standard stars and solved the transformation
equations between the instrumental and standard systems. We
obtained
\begin{eqnarray}
m_{std,V} & = & m_{obs,V}  + const_V - 0.161 X - 0.065 (V-I), \label{eq_mv} \\
m_{std,I} & = & m_{obs,I}  + const_I - 0.071 X - 0.039 (V-I)  \label{eq_mi}
\end{eqnarray}
where $m_{std}$ is the magnitude of the standard star, $m_{obs}$ is the actually
observed magnitude, $X$ is the airmass and  $(V-I)$ is the color of the star.
The residuals were 0.010 $mag$ in $V$ and 0.011 $mag$ in $I$.
The extinction coefficients are close to the average values found for SSO
\citep{sb00} and the color terms are virtually identical to those found by
\citet{sbl98} for the particular camera and filter set that we used.

The photometry for each Fornax frame was extracted using DoPHOT \citep{sms93}
in fixed position mode.
We first created a ``template'' frame for each field. Each template is the
average of 4-6 good seeing images. Photometry was done on this deep template
and the star list of the template was used as input for the photometry of each
program frame. Then all images were put on the same zero-point (still in
the instrumental system). We applied an aperture correction to these magnitudes
and Eqs~(\ref{eq_mv}) and (\ref{eq_mi}) were inverted
to provide calibrated photometry for each frame. We estimate that our absolute
zero point is accurate to $\sim 0.02\ mag$.

\subsection{Observations at Mount Stromlo Observatory}

We obtained 20 images with the wide-field camera mounted on the Mount
Stromlo Observatory 50'' telescope (MSO).
This camera was used by the MACHO project and is a mosaic of
four 2k$\times$2k CCD with a pixel scale of 0.628''/pixel,
covering approximately $42' \times 42'$. A feature
of this camera is a dichroic mirror that allows to obtain simultaneous
measurements in two passbands, $V_M$ and $R_M$,
corresponding (very roughly) to $V$ and $R$.
We refer the reader to \cite{st93} and \cite{a99} for descriptions of the
camera system and filters. The exposure time was 10 minutes for each image.
The typical seeing was 2.2''.

Photometry was also performed in fixed-position mode and the frames were then put
on a common zero point. Since the passbands are non-standard, we needed to calibrate
$V_M$ and $R_M$ to $V$ and $I$. For this
we could use our already calibrated photometry from the 1m SSO
telescope. The passbands $V_M$ and $R_M$ are quite broad but they have been
calibrated to $V$ and $R$ (Bessell \& Germany 1999, Alcock et al. 1999).

The areas covered with each telescope/instrument combination are
virtually identical, thus we have $VI$ and $V_M R_M$ photometry for most stars.
Figure~\ref{fig_vrmvi} illustrates the procedure we used to calibrate the color
$V_M-R_M$ to $V-I$. 
For each chip, we constructed plots like Fig.~\ref{fig_vrmvi} and plots of $I-R_M$
vs $V_M-R_M$. The ``knee'' at $V_M-R_M \simeq 1.3$ can already be found
in \cite{bg99} in a $V-R_M$ vs $V_M - R_M$ diagram (their Fig.~5).
This knee also happens
when we use $V-I$ since there is an almost linear relation between $V-I$ and $V-R$.
We then made separate fits for stars bluer and redder than $V_M-R_M = 1.3$. 
For blue stars ($V_M-R_M\leq 1.3$) a typical calibration would be 

\begin{eqnarray}
V-I   & = & 1.123 (V_M-R_M)  + 0.069 \label{eq_vicalblue}\\
I-R_M & = & -0.341 (V_M-R_M) - 0.141 \label{eq_icalblue}
\end{eqnarray}

and for red stars ($V_M-R_M > 1.3$)

\begin{eqnarray}
V-I   & = & 2.024 (V_M-R_M)  - 1.083 \label{eq_vicalred}\\
I-R_M & = & -0.842 (V_M-R_M) + 0.502 \label{eq_icalred}
\end{eqnarray}

From chip to chip there were slight differences in the coefficients, amounting
to a few percents at most. Our separate chip-by-chip calibration might not have been
warranted (the approach followed by Bessell \& Germany 1999).  However, we
certainly do not lose accuracy in the process, it is only a little more cumbersome.


After the calibration of the whole data set
an astrometric calibration was done, using the USNO-A2.0 catalog \citep{usno} giving
residuals between 0.3 and 0.5 arcseconds.

With calibrated photometry in hand, the last step is to apply a reddening
correction. Given the high Galactic latitude of Fornax ($b \sim -66^\circ$),
the average reddening should be very low. We thus used the maps of \citet{sfd98}
to determine the reddening to every star in our data set. The average $E(B-V)$
is around 0.025 mag.

\section{Search for variables}

We used the method proposed by \cite{ws93}. It makes use of the fact that when
a star is brighter (or fainter) than average in one passband it is also brighter
(fainter) than average in another passband. The variability index $I_{WS}$ has
been computed for each star, and Fig~\ref{fig_iws} presents the results.
Fig~\ref{fig_cmdvar} shows the color-magnitude diagram with the variables
we have identified.

%
%

Fornax has five globular clusters \citep{ho61}, two of which are in our field
(cluster 2 and cluster 4). However it was impossible to find variables in these
clusters, because of the exceedingly high density of stars. While it is possible
to resolve some stars in cluster 2 (no star could be resolved in cluster 4),
the effect of a varying seeing made the photometry extremely unreliable,
although we found one candidate Long Period Variable in cluster 2.
Detection of RR Lyrae in Fornax's globular clusters can only be done with
a much better spatial resolution than we have.


The probability of detecting a variable star depends on a number of
things, an important one being the actual time distribution of the
observations, since this can affect the detection of some periods more
than others.  We carried out numerical simulations (much in the spirit
of Saha \& Hoessel, 1990) to estimate the impact that our sampling may
have on the discovery of variable stars. For a star of period $P$ and
initial phase $\phi_0$ (phase at the time of the first observation),
we can compute the phase at each of our observations. We then
determine which of the following criteria are fulfilled:
\begin{enumerate}
  \item At least 3 observations between phases 0 and 0.2 ($\phi=0$
     corresponds to maximum light)
  \item At least 2 phases between 0.2 and 0.5 (descending light)
  \item At least 3 phases between 0.5 and 0.8 (around minimum light)
  \item At least 15 measurements in total
\end{enumerate}

For all periods between 0.1 and 50 days, and for all initial phases,
we simply count the number of observations satisfying each criterion,
after having attributed a weight to each observation: we estimate that
the \emph{photometric} detection efficiency of the SSO 40'' data is
0.8, and 0.4 for the MSO 50'' data. This means that, for instance,
four actual SSO 40'' observations are nedded to satisfy criterion 1.
For each period we then count the number of phases that fulfill these
criteria. After appropriate normalization, we obtain a period spectrum
which is simply the probability of detecting a star as a function of
period (see Fig~\ref{fig_effic}).

\subsection{RR Lyrae stars}

We selected all stars in a broad region around the horizontal branch
in order to look for RR Lyrae variables. Specifically the criteria
were $20.8 \leq V \leq 21.6$, $0.2 \leq V-I \leq 0.8$ and $I_{WS} >
0.9$.  Our data are somewhat inhomogeneous in the sense that that SSO
40'' data are of better quality than the MSO 50'' data. In the search
for RR Lyrae we thus used only SSO data when computing the $I_{WS}$
index. The value $I_{WS}=0.9$ has been chosen as a compromise between
finding the largest number of genuine variables while minimizing the
number of false detections. It can be seen from Fig~\ref{fig_iws} that
the bulk of the stars are below $I_{WS}=0.9$.
Applying the criteria given above yielded 851 objects.

We used the phase-dispersion minimization algorithm (Stellingwerf,
1978) to look for periodicity in $V$-band data. A rapid inspection of
some light curves revealed that for a fairly large number of
candidates no period could give a reasonable light curve. This is
because the magnitude of a RR Lyrae variable is close to the detection
limit of our photometry.  We thus decided to apply a conservative
approach and to select the best period by examining each light curve.
This led to rejecting a fairly large number of candidates.  Our goal
was to produce a reasonably clean sample of RR Lyrae, not a complete
one. Among the stars rejected, a significant number may be genuine RR
Lyrae. 

The main reason to reject a candidate variable was the total number of
measurements per star. Among the rejected candidates, 48\% had less
than 25 measurements whereas 75\% of the confirmed RR Lyrae have more
than 25 data points.  
There is no obvious correlation between the value of the variability
index $I_{WS}$ and the probability of finding a period. This is
because we used $\sim 15$ epochs to construct the index $I_{WS}$ (only
SSO 40'' data). Some stars may show a large $I_{WS}$ but may have only
15 measurements, making it difficult to find a reliable period.
The second reason for rejecting a variable is
aliasing. For some stars it was impossible to decide between two
periods. In particular, for a given frequency $\nu$, the light curves
for $1-\nu$ or $1+\nu$ would be almost as convincing as $\nu$. In the
end, this selection yielded 515 RR Lyrae stars for which one single
frequency could be clearly identified, hence 60\% of our candidates
turned out to be ``good'' RR Lyrae variables. They are listed in
Table~\ref{tbl_rrl} where we give the phase-weighted average
magnitudes and associated uncertainties. Some light curves are shown
in Fig.~\ref{fig_lcrr}.  When considering the $I$-band light curves in
this figure, it is quite remarkable that the Welch-Stetson method
still works, even though the variability signal is extremely weak in
the $I$ band. This shows that the method is very powerful even at
faint magnitudes.  One has to bear in mind that the number of
variables is likely to be a lower limit to the total number of RR
Lyrae stars in Fornax.  In particular, a consequence of the quality of
our data (the average measurement error for a typical RR Lyrae
variable is $\sim 0.15\ mag$) is that our survey is probably biased
against RRc variables, which have a low amplitude. This will affect
the relative number of RRab and RRc variables but the incompleteness
should not depend on period.  With better photometric data it will be
possible to obtain good light curves for many more RR Lyrae.


The period histogram is shown on Fig.~\ref{fig_rrhistop}. There are two
peaks, one at $\sim 0.57$ days and one at $\sim 0.38$ days.  There
seems to be a gap between $0.46^d$ and $0.48^d$. The probability of
detecting variables with periods around $0.5^d$ is very low but for
periods $0.46^d \leq P \leq 0.48^d$ our detection probability is
always larger than 66\% of the maximum (see Fig.~\ref{fig_rrhistop})
so we should be able to discover variables with periods between
$0.46^d$ and $0.48^d$, if they exist.  We thus interpret this gap as
the transition period from RRab to RRc. The shortest period for RRab
type is thus $0.48^d$.  Given that our data are not good enough to
determine the type of variable on the basis of the light curve, we
\emph{defined} RRab type variables as the stars having periods larger
than 0.47 days, and RRc as having periods shorter than 0.47 days. The
number of stars in each category is $N_{ab} = 396$ and $N_c = 119$. We
then determined the average periods $<P_{ab}> = 0.585^d$ and $<P_{c}>
= 0.349^d$. The longest period for RRab variables is 0.737$^d$.


In galactic globular clusters, the average period of ab type RR Lyrae
variables $<P_{ab}>$ is either near $0.55^d$ or near $0.64^d$
(Oosterhoff class I and II respectively).  The surprising fact is that
dwarf spheroidal galaxies do not fall in either of these
categories. From Table~7 in \citet{mfk95}, one sees that all dwarf
spheroidal galaxies are intermediate between Oo I and Oo II
categories.  Adding the result of \citet{sm00} for Leo II ($<P_{ab}> =
0.619^d$) and the present value of $0.585^d$ for Fornax only
reinforces this statement.

The average properties of RR Lyrae variables are related to their metallicity.
There are relations between several characteristic periods (longest, shortest, average)
and the metal content \citep{s93a}. For the shortest period one has
\begin{eqnarray}
\log P_{ab} = -0.122 (\mbox{[Fe/H]}) - 0.500 \label{eq_logpab}
\end{eqnarray}
which yields $\mbox{[Fe/H]} = -1.5$ (for $P_{ab}=0.48^d$).  
Through the relation
\begin{eqnarray}
\log < P_{ab} > = -0.092 (\mbox{[Fe/H]}) - 0.389 \label{eq_logp}
\end{eqnarray}
the average period yields $\mbox{[Fe/H]} = -1.7$. The longest period
is related to [Fe/H] via
\begin{eqnarray}
\log P_{ab} = -0.09 (\mbox{[Fe/H]}) - 0.280 \label{eq_logplong}
\end{eqnarray}
which gives $\mbox{[Fe/H]} = -1.64$.
All these values of [Fe/H] are on the Butler-Blanco
metallicity scale (used by Sandage) which is more metal-rich than the
widely used scale of \citet{zw84} by 0.2 dex.

When considering in particular the average period $<P_{ab}>$, there
is some scatter in this relation, about 0.2 dex (see Siegel \&
Majewski 2000, their figure 6, which also displays the other eight
dwarf spheroidal galaxies). Our estimates of the metal abundance
cluster around $\approx -1.6$, we will then take as a final value
$\mbox{[Fe/H]} = -1.6 \pm 0.2$.

In contrast with this small dispersion in [Fe/H] is the fairly large
dispersion in average magnitudes for RR Lyrae stars ($\sigma = 0.140$).
This is significantly larger than what is observed in other dwarf
spheroidal galaxies [e.g. $\sigma = 0.096$ in Leo II -- \citet{sm00},
$\sigma = 0.104$ in Sculptor -- \citet{k95}]. It seems likely that
photometric scatter is reponsible for the dispersion in $<V>$.

\subsection{Population II and anomalous Cepheids}

We found a number of variables slightly or significantly brighter than
the horizontal branch, most of which are anomalous Cepheids (ACs) or
Population II Cepheids. Their magnitudes and periods are given in
Table~\ref{tab_acs}.  We found 6 Pop II Cepheids (one unsure), 17
anomalous Cepheids (one unsure) and one blue variable whose nature is
unclear.


The position of a variable in the period-luminosity diagram
(Fig.~\ref{fig_pl}) is determined by its metallicity and pulsation
mode (e.g. Nemec et al 1994).  Fornax has a large range in [Fe/H] and
it has had a complex star formation history \citep{shb00}. This means
that Cepheids in this galaxy can have a range of ages and/or
metallicities. Unfortunately, in the absence of any information on
[Fe/H] it is impossible to assign a pulsation mode to these
variables. By looking at Fig~\ref{fig_pl} it is clear however that
there must be a range of metallicity among these stars.  Some of the
extreme stars in Fig.~\ref{fig_pl} could be metal-poor and overtone
pulsators while others are metal-rich and fundamental pulsators.


There is a surprising correlation between the specific frequency of
anomalous Cepheids and the luminosity of the dwarf spheroidal galaxies
\citep{mfk95}. According to this relation, about a dozen anomalous
Cepheids should have been found. We found 17, fairly close to the
prediction and not large (nor small) enough to destroy the correlation
found by \citet{mfk95}.  The nature of anomalous Cepheids is
unclear. They are found in systems where the most massive stars still
burning nuclear fuel are $\approx 0.8 M_\odot$, however the masses of
anomalous Cepheids are around $\sim 1.5 M_\odot$ (e.g. Wallerstein \&
Cox 1984, Bono et al. 1997) The two explanations usually put forward
to explain their large masses are that they are intermediate-age stars
or that they are the product of a binary star merging (the two
component having the mass of a turn-off star).  Fornax has an
important intermediate-age population so the first hypothesis is
perfectly plausible, however it is not easy to rule out the second
possibility.

A remarkable fact is that, among the nine dwarf galaxies satellites of
the Milky Way, Fornax is the only one containing Pop. II
Cepheids. These stars are in a short-lived evolutionary stage and
Fornax is probably the only dwarf galaxy that is massive enough to
have a few of these stars.

\subsection{Long Period Variables}

There has already been a wide-field survey for Long Period Variables
(LPVs) in Fornax \citep{di87}. They covered the whole galaxy and found
30 variables.  It was based on photographic plates obtained at the UK
Schmidt telescope and they acknowledge the fact that ``a certain
number of bright variables must have been missed ... because of
blended images''. It thus seemed worthwhile to look specifically for
bright red variables showing a long-term trend, even though our
sampling is far from ideal for these stars.  \citet{dk79} presented a
list of very red stars in Fornax. Based on two $V$ measurements
separated by 132 days, they could flag some of these stars as
candidate variables. We also tried to recover objects from that paper.

Eighty five stars turn out to be reasonably good candidates. They are
given in Table~\ref{tbl_lpv}.  In this table, a question mark means
that the variability is not certain; this happens particularly for
crowded stars. Given that our data do not allow to obtain periods for
such objects, the classification as LPVs is only tentative.  However,
given their magnitudes and colors, most objects in Table~\ref{tbl_lpv}
are good candidates.  In this Table we also indicate the numbers given
by \citet{shsh} in their Table~1 for red stars.

We have data for 19 variables listed by \citet{di87}, 3 of which
appear non variable in our data simply because we don't have enough
measurements. The remaining 11 stars in the list of Demers \& Irwin
are not in our survey area.  In addition we find 38 good candidate
LPVs and 28 possible variables (indicated with a ? in
Table~\ref{tbl_lpv}). In Fig.~\ref{fig_lpv} we show two examples of
LPVs.


There are a few dozens of red giants ($V<19,\ V-I>1$) showing some
variability (i.e.  with $I_{WS} > 0.9$), most with small amplitude
(less than 0.1 $mag$).
Variability along the red giant branch has been
detected in a number of cases in Galactic field and globular clusters
giants (e.g. Smith \& Dupree 1988 and references therein), however it
is not clear whether this is related to pulsation, spots, or any other
mechanism.  The typical periods of these objects are likely to be in
the range of days to weeks, where our ability to find a period is
seriously affected by the sampling (see Fig~\ref{fig_effic}).

\section{Discussion}

We have presented a survey of $1/2$ square degree covering the central
region of the Fornax dwarf galaxy. We found and determined periods for
17 anomalous Cepheids, 6 Population II Cepheids and more than 500 RR
Lyrae. In addition, 85 LPV candidates were identified.  It is almost
certain that there are more RR Lyrae in this galaxy: the accuracy of
our data do not allow us to produce a complete catalog of RR
Lyrae. The average metal abundance of RR Lyrae is $\mbox{[Fe/H]}_{RR}
\simeq -1.6 \pm 0.2$.

We can use the known distance to Fornax ($\mu_0 = 20.68$ with the tip
of the red giant branch, Bersier 2000) to estimate the brightness of
RR Lyrae. With an average magnitude of $<V_0> = 21.27$, one obtains
$M_V(RR) = 0.59 \pm 0.1\ mag$. This is intermediate between other estimates
of the magnitude of RR Lyrae stars. \citet{gp98} found $M_V(RR) \simeq
0.75$ for [Fe/H] = -1.7, while from \citet{s93b} one would get
$M_V = 0.40\ mag$.

\acknowledgments

We thank the director of Mt Stromlo \& Siding Spring Observatories,
Jeremy Mould, for the allocation of director's discretionary time on
the MSO 50'' telescope.  We also thank the referee for a swift and
constructive report that helped us improve this paper.  DB
acknowledges partial support from the Swiss National Science
Foundation (grant 8220-050332). This work has also been supported by
NSF grant AST-9979812.

\clearpage



\begin{table}
\begin{center}
\caption{Journal of observations.}
\label{tbl_dates}
\begin{tabular}{ l l l l }
\tableline\tableline
Field & JD$_V$ & JD$_I$ & Observatory \\
\tableline
1 & 652.26667 & 652.26667 & MSO \\
1 & 657.26976 & 657.26976 & MSO \\
1 & 658.27549 & 658.27549 & MSO \\
1 & 659.28311 & 659.28311 & MSO \\
1 & 660.25075 & 660.25075 & MSO \\
1 & 661.23904 & 661.23904 & MSO \\
1 & 662.23573 & 662.23573 & MSO \\
1 & 663.23345 & 663.23345 & MSO \\
1 & 664.23512 & 664.23512 & MSO \\
\tableline
\end{tabular}

\tablecomments{Table~\ref{tbl_dates} is presented in its complete form in the
electronic version of the journal. Only a fraction is shown here for guidance
regarding its form and content.}

\end{center}
\end{table}

\begin{table}
\begin{center}
\caption{RR Lyrae variables.}
\label{tbl_rrl}
\begin{tabular}{ l r c c r c c r c c }
\tableline\tableline
 Name  & $N_V$ & $V$ & $\sigma_V$ & $N_I$ & $I$ & $\sigma_I$ & $I_{WS}$ & Period (days) & $E(B-V)$ \\
\tableline
FBW J023749.8-342427 & 24 & 21.389 & 0.027 & 24 & 20.755 & 0.036 & 1.422 & 0.54365 & 0.028 \\
FBW J023750.8-342736 & 32 & 21.346 & 0.021 & 31 & 20.744 & 0.035 & 2.741 & 0.58252 & 0.030 \\
FBW J023751.9-342620 & 32 & 21.419 & 0.023 & 32 & 20.887 & 0.033 & 1.139 & 0.58052 & 0.029 \\
FBW J023752.8-344615 & 32 & 21.469 & 0.029 & 32 & 20.560 & 0.033 & 2.005 & 0.16459 & 0.034 \\
FBW J023753.4-345215 & 28 & 21.344 & 0.026 & 28 & 20.781 & 0.035 & 1.584 & 0.59260 & 0.035 \\
FBW J023755.0-343726 & 30 & 21.401 & 0.026 & 29 & 20.911 & 0.035 & 1.506 & 0.40322 & 0.035 \\
FBW J023756.0-342732 & 32 & 21.468 & 0.020 & 32 & 21.101 & 0.036 & 5.965 & 0.33797 & 0.030 \\
FBW J023756.1-344332 & 31 & 21.544 & 0.022 & 31 & 20.849 & 0.036 & 2.666 & 0.51202 & 0.034 \\
FBW J023756.6-343507 & 32 & 21.392 & 0.022 & 32 & 20.766 & 0.031 & 2.208 & 0.54362 & 0.035 \\
FBW J023757.0-342201 & 32 & 21.455 & 0.021 & 32 & 20.838 & 0.032 & 2.497 & 0.56046 & 0.028 \\
\tableline
\end{tabular}

\tablecomments{Table~\ref{tbl_rrl} is presented in its complete form in the
electronic version of the journal. Only a fraction is shown here for guidance
regarding its form and content.}

\end{center}
\end{table}

\clearpage

\begin{table}
\begin{center}
\caption{Anomalous Cepheids and Population II Cepheids
\label{tab_acs}}
\begin{tabular}{ l c c c c c r l }
\tableline\tableline
 Name    & $V$ & $V-I$ & $\sigma_V$ & $\sigma_I$ & $E(B-V)$ & Period & Class \\
\tableline
 FBW J023803.9-343822 & 19.478 & 0.483 & 0.007 & 0.012 & 0.036 & 15.2283 & P2C  \\ 
 FBW J023811.9-344023 & 20.651 & 1.087 & 0.015 & 0.021 & 0.036 & 1.21327 & P2C  \\ 
 FBW J023843.0-344825 & 19.801 & 0.558 & 0.016 & 0.031 & 0.031 & 1.04503 & AC   \\ 
 FBW J023852.4-343012 & 19.905 & 0.602 & 0.007 & 0.011 & 0.027 & 1.24968 & AC   \\ 
 FBW J023853.4-343048 & 19.712 & 0.564 & 0.006 & 0.010 & 0.027 & 2.57976 & P2C  \\ 
 FBW J023907.1-343316 & 20.786 & 0.774 & 0.017 & 0.025 & 0.026 & 0.50812 & AC   \\ 
 FBW J023926.9-342806 & 20.272 & 0.667 & 0.010 & 0.016 & 0.025 & 2.08323 & P2C  \\ 
 FBW J023926.8-343122 & 20.828 & 0.673 & 0.017 & 0.028 & 0.024 & 0.50398 & AC   \\ 
 FBW J023927.0-342427 & 21.052 & 0.082 & 0.019 & 0.042 & 0.026 & 0.56965 & blue \\ 
 FBW J023937.7-343621 & 20.645 & 0.725 & 0.015 & 0.024 & 0.025 & 0.546   & AC   \\ 
 FBW J023941.5-343125 & 20.616 & 0.475 & 0.012 & 0.022 & 0.024 & 0.57345 & AC   \\ 
 FBW J023945.2-344314 & 21.230 & 0.481 & 0.019 & 0.037 & 0.024 & 1.14259 & P2C? \\ 
 FBW J023946.2-343001 & 20.568 & 0.786 & 0.014 & 0.023 & 0.023 & 0.92161 & AC   \\ 
 FBW J023952.5-343340 & 20.252 & 0.904 & 0.014 & 0.023 & 0.024 & 1.31071 & AC   \\ 
 FBW J023953.5-342433 & 20.846 & 0.656 & 0.015 & 0.026 & 0.022 & 0.61144 & AC   \\ 
 FBW J023954.8-343601 & 21.022 & 0.478 & 0.016 & 0.031 & 0.025 & 0.57442 & AC   \\ 
 FBW J024000.9-341846 & 20.828 & 0.476 & 0.015 & 0.025 & 0.023 & 0.41592 & AC   \\ 
 FBW J024002.7-341928 & 20.362 & 0.849 & 0.009 & 0.015 & 0.022 & 0.53299 & AC   \\ 
 FBW J024010.2-343212 & 20.624 & 0.412 & 0.013 & 0.025 & 0.022 & 1.33531 & P2C  \\ 
 FBW J024016.9-343641 & 20.840 & 0.430 & 0.014 & 0.030 & 0.025 & 0.53294 & AC   \\ 
 FBW J024017.4-342024 & 19.994 & 0.608 & 0.007 & 0.011 & 0.021 & 1.19847 & AC   \\ 
 FBW J024022.8-342802 & 20.480 & 0.526 & 0.011 & 0.019 & 0.020 & 0.83757 & AC   \\ 
 FBW J024050.2-344335 & 20.826 & 0.729 & 0.015 & 0.024 & 0.024 & 0.50578 & AC   \\ 
 FBW J024058.3-344552 & 20.786 & 0.663 & 0.018 & 0.028 & 0.024 & 0.48106 & AC   \\ 
\tableline
\end{tabular}
\end{center}
\end{table}

\clearpage

\begin{table}
\begin{center}
\caption{Candidate and known Long Period Variables.}
\label{tbl_lpv}
\begin{tabular}{ l r c c r c c c l l }
\tableline\tableline
 Name & $N_V$ & $V\tablenotemark{a}$ & $\sigma_V\tablenotemark{b}$ & $N_I$ & $V-I\tablenotemark{a}$ & $\sigma_{V-I}\tablenotemark{b}$ & $E(B-V)$ & Other ID\tablenotemark{c} & Note\tablenotemark{d} \\
\tableline
FBWJ023800.0-344422 & 32 & 18.562 & 0.021 & 32 & 1.860 & 0.026 & 0.034 &            &   \\
FBWJ023806.1-343119 & 30 & 20.233 & 0.065 & 30 & 3.765 & 0.070 & 0.033 &            &   \\
FBWJ023812.6-345606 & 32 & 18.504 & 0.022 & 32 & 2.291 & 0.026 & 0.033 & DK1, DI30  &   \\
FBWJ023817.1-342548 & 32 & 18.591 & 0.023 & 32 & 2.314 & 0.030 & 0.029 & S3, DI13   &   \\
FBWJ023821.3-343618 & 32 & 18.335 & 0.021 & 32 & 1.942 & 0.026 & 0.034 &            & ? \\
FBWJ023822.0-345213 & 32 & 17.960 & 0.014 & 32 & 1.577 & 0.019 & 0.032 &            &   \\
FBWJ023822.6-343804 & 32 & 18.731 & 0.023 & 31 & 2.296 & 0.029 & 0.035 & DK15       &   \\
FBWJ023826.3-342533 & 30 & 18.451 & 0.017 & 30 & 1.649 & 0.022 & 0.029 &            &   \\
FBWJ023830.2-344503 & 28 & 18.324 & 0.017 & 28 & 1.875 & 0.021 & 0.033 &            &   \\
FBWJ023833.4-342824 & 32 & 18.403 & 0.018 & 32 & 1.717 & 0.025 & 0.029 &            & ? \\
\tableline
\end{tabular}

\tablenotetext{a}{The magnitudes are arithmetic averages.}
\tablenotetext{b}{The $\sigma$s are the average measurement errors.}
\tablenotetext{c}{Number preceded by S are from Table~1 of Stetson et al. (1998),
numbers preceded by DK are from Demers \& Kunkel (1979), numbers preceded by
DI are from Demers \& Irwin (1987).}
\tablenotetext{d}{A ? means that the variability is not certain.}

\tablecomments{Table~\ref{tbl_lpv} is presented in its complete form in the
electronic version of the journal. Only a fraction is shown here for guidance
regarding its form and content.}

\end{center}
\end{table}

\clearpage



\figcaption{
The color $(V-I)$ as a function of MACHO color $(V_M-R_M)$.
The two segments of straight lines show the calibration applied to the data
\label{fig_vrmvi}}

\figcaption{
The variability index $I_{WS}$ as a function of magnitude.
Stars with $I_{WS} \geq 0.9$ have a large probability of being variable.
The vertical ``finger'' at $m_V \simeq 21.3$ is caused by RR Lyrae.
\label{fig_iws}}

\figcaption{
The color-magnitude diagram. Note that the magnitudes of non-variable stars are the
average of 12 measurements. Crosses are for RR Lyrae, triangles are for anomalous 
and Population II Cepheids, hexagons are for Long Period Variables.
\label{fig_cmdvar}}

\figcaption{
The probability of detecting variable stars as a function of period
for field 1 (see text for explanations).
{\it Upper panel:} for periods between $0.1^d$ and $50^d$,
{\it Lower panel:} for periods between $0.1^d$ and $1^d$.
\label{fig_effic}}

\figcaption{
$V$ and $I$ light curves of six RR Lyrae variables, including the longest period one
(bottom right).
The period is indicated for each star in the upper left part of the $V$ panel.
\label{fig_lcrr}}

\figcaption{
({\it Solid line}) The period histogram for all confirmed RR
Lyrae. The absence of variables with periods near 0.46 days suggest
that this is the transition period between RRab and RRc type
variables.
({\it Dashed line}) The detection probability for field 2
\label{fig_rrhistop}}

\figcaption{
The top panels show the $V$ and $I$ light curves of two anomalous Cepheids;
the bottom panels show the light curves of two Pop II Cepheids.
The periods are indicated in each case.
\label{fig_acs}}

\figcaption{
The period-luminosity plot for anomalous Cepheids (squares) and
Population II Cepheids (stars). The solid lines are the PL relations
for fundamental and first overtone modes from Nemec et al. (1994) with
[Fe/H]$=-2$, shifted to correspond to a distance modulus of 20.66 mag
\citep{db00}; the dotted lines are the same for [Fe/H]$=-1$.  The
cross indicates the average position of RR Lyrae stars.
\label{fig_pl}}

\figcaption{
$V$ and $I$ data for two candidate Long Period Variables. The top one
is DI13, the second has not been found by \citet{di87}. The triangles
are for $I$-band data, squares are for $V$-band data.
\label{fig_lpv}}

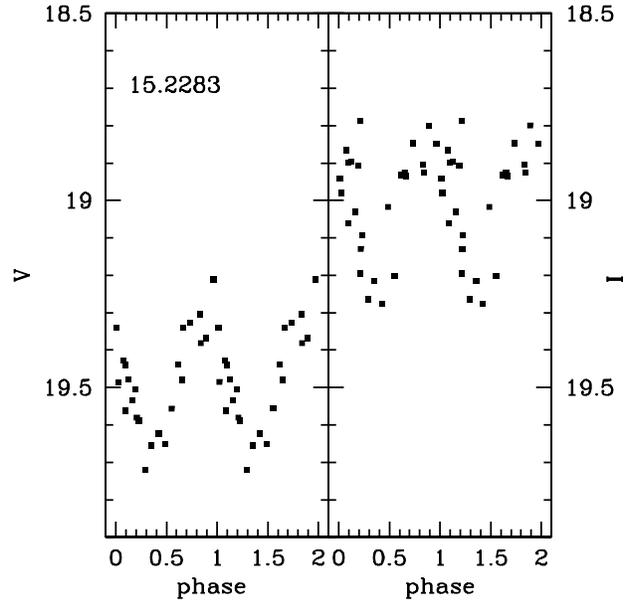
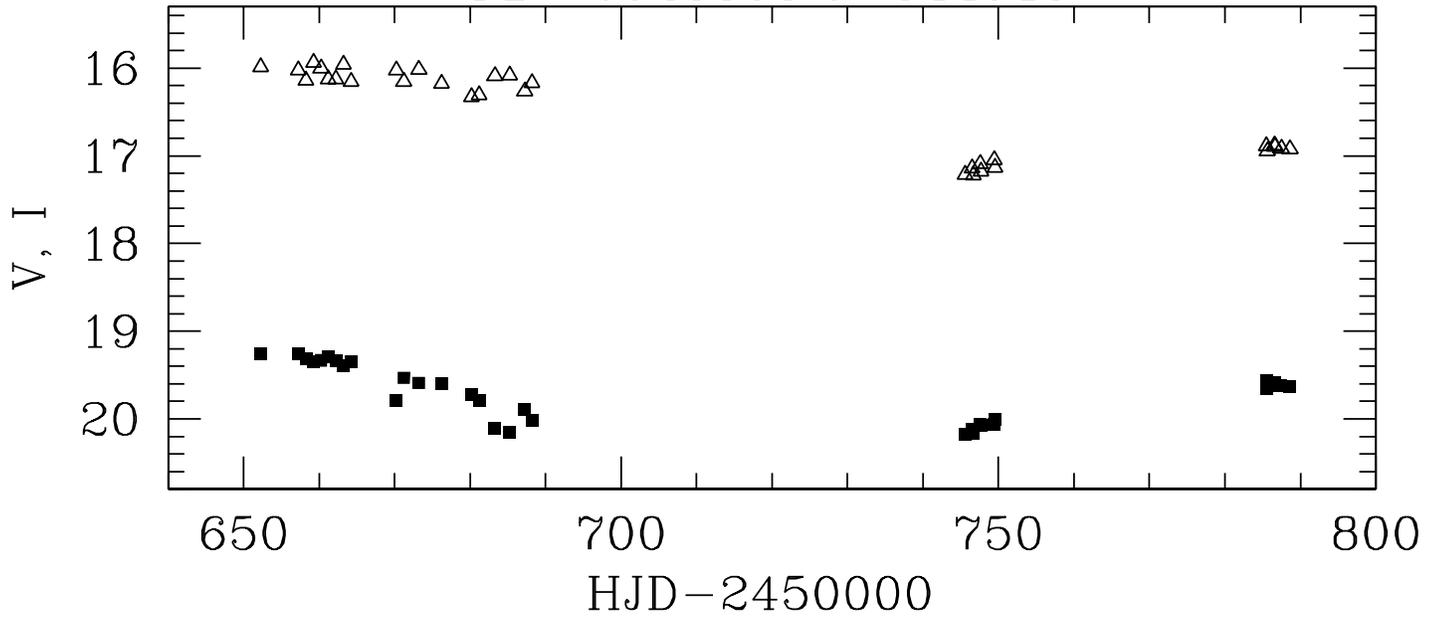
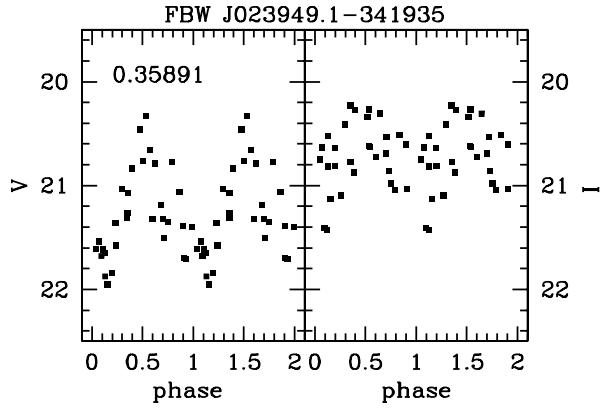
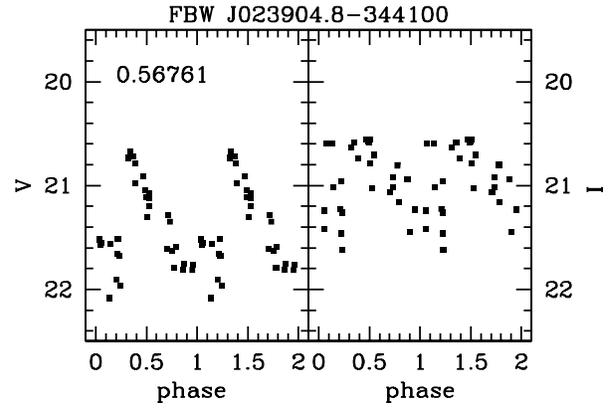
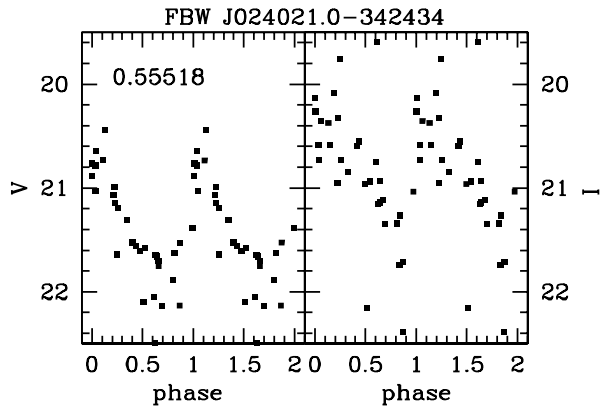
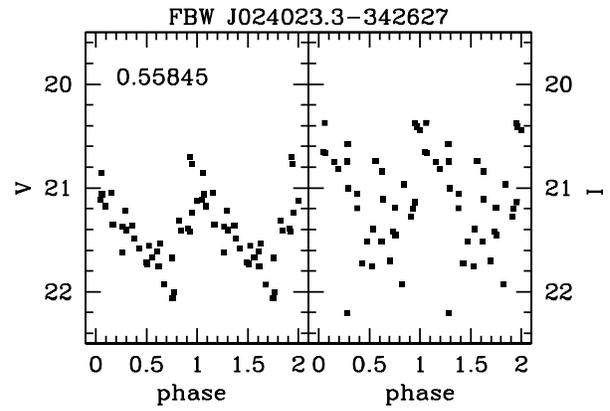
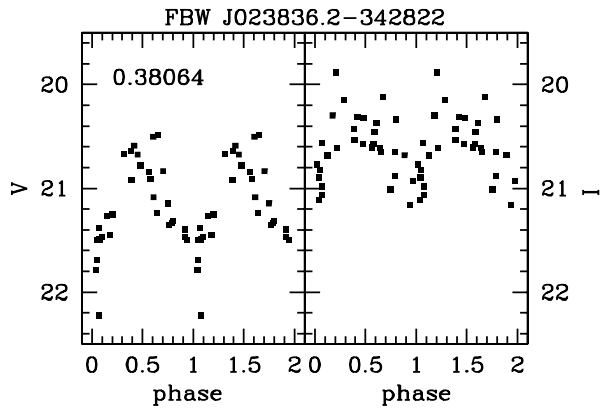
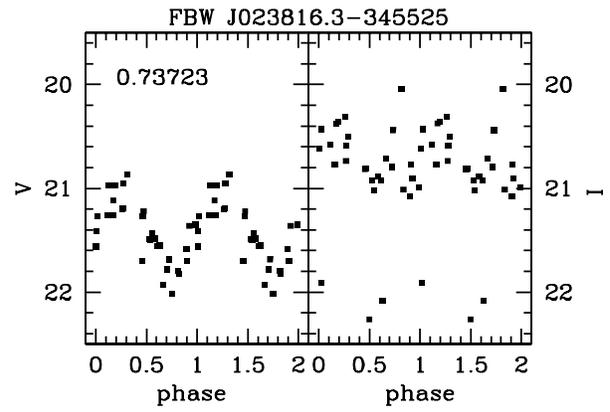
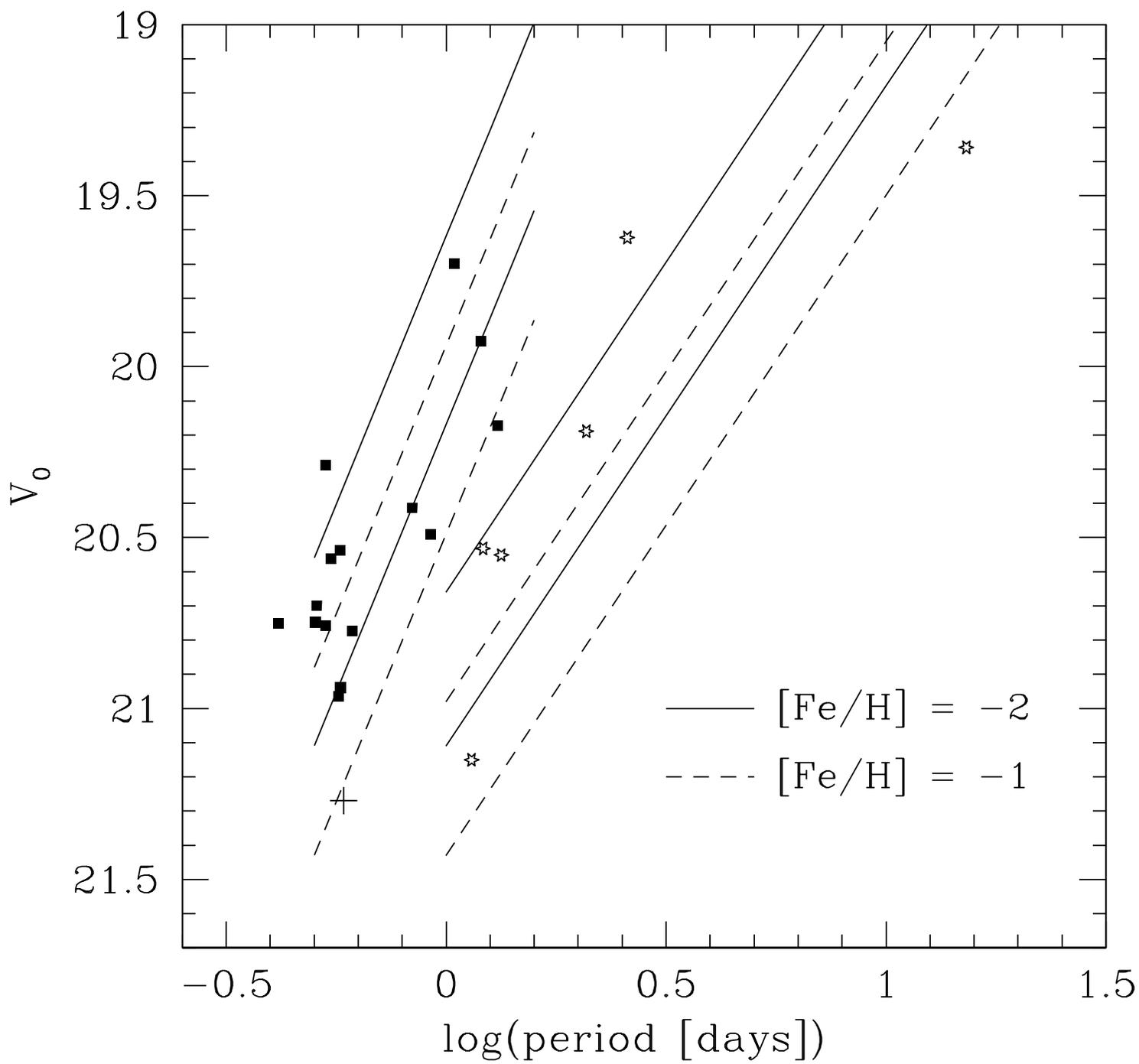
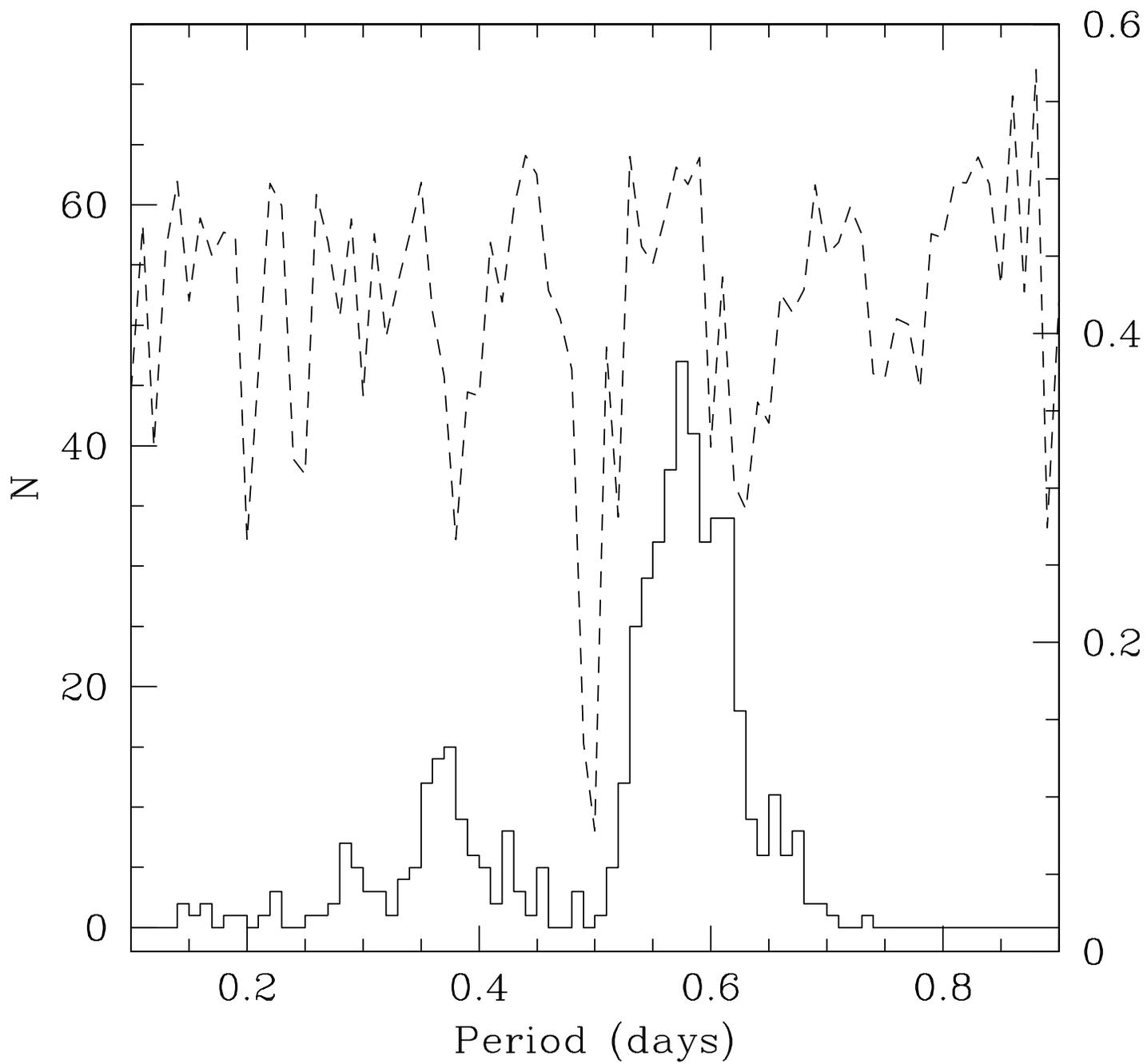
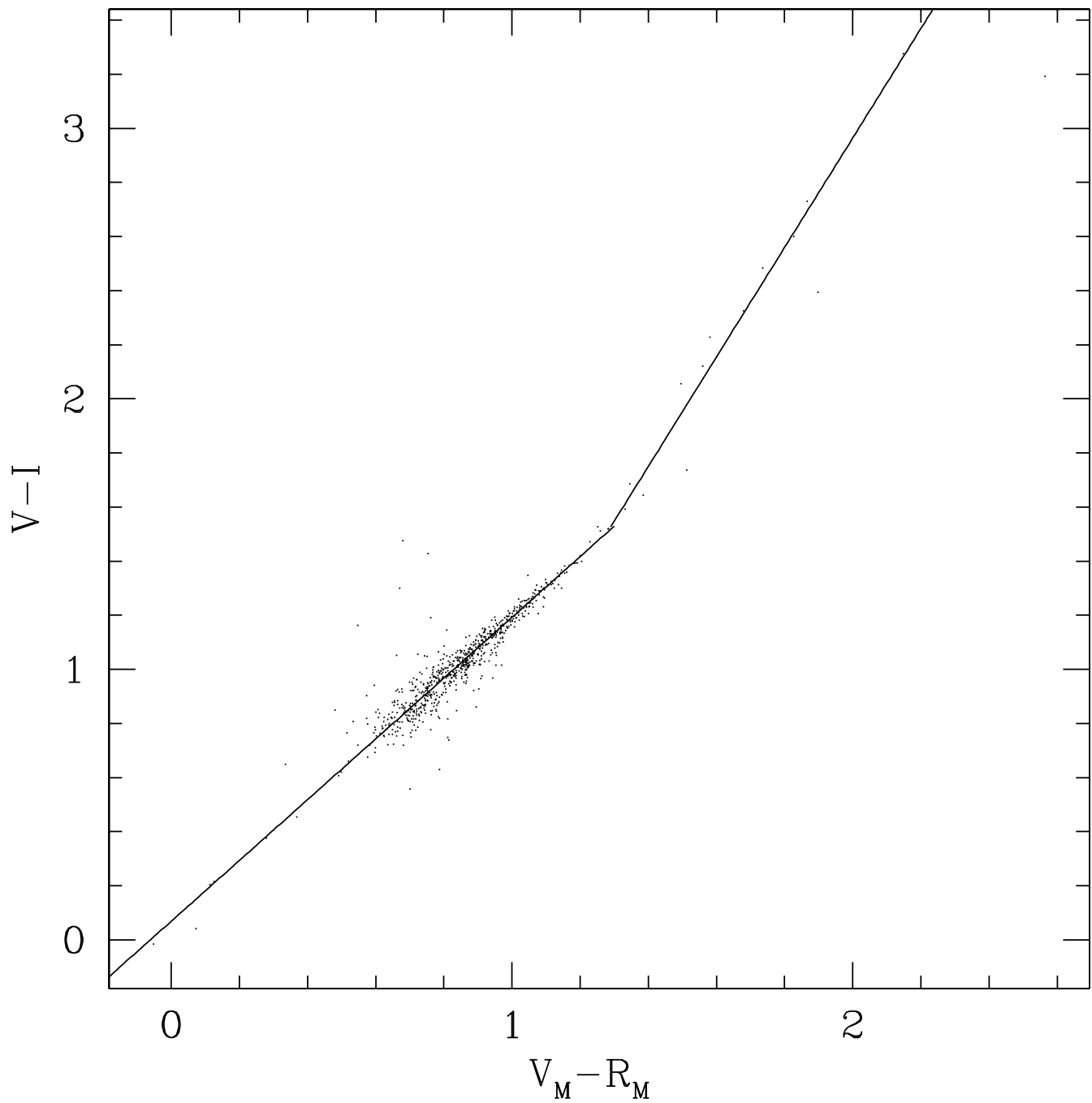
\end{document}